\documentclass[12pt]{article}

\usepackage{graphicx,a4}
\usepackage{amssymb}
\usepackage{latexsym,epsfig}
\begin{document}
\newcommand{\preprintno}[1]
{\vspace{-2cm}{\normalsize\begin{flushright}#1\end{flushright}}\vspace{1cm}}\title{\preprintno{{\bf ULB-TH/04-06}}Cosmic acceleration in models with density dependent moduli}
\author{{Malcolm Fairbairn\thanks{E-mail: mfairbai@ulb.ac.be}}\\
{\em Service de Physique Th\'eorique, CP225}\\
{\em Universit\'e Libre de Bruxelles, B-1050 Brussels, Belgium}}\date{May 2004}
\maketitle
\begin{abstract}
The effective equation of state of normal matter is changed in theories where the size of the compact space depends upon the local energy density.  In particular we show how the dilution of a fluid due to the expansion of the universe can be compensated by an increase of the effective coupling of that fluid to gravity in the presence of a potential which acts to reduce the size of the compact space.  We estimate how much cosmic acceleration can be obtained in such a model and comment on the difficulties faced in finding an appropriate potential.
\end{abstract}
\section{Introduction}
Current astrophysical observations suggest that there are at least two
separate epochs in the history of the universe where accelerated
expansion occurred.  The first is in the very early universe where a
period of accelerated expansion could explain the uniform temperature
of the cosmic microwave background radiation across the sky.  The
second is the apparent acceleration of the universe deduced from
observations of type 1a supernovae which seems to be occurring today
and to have begun extremely recently in cosmological terms
\cite{sn1a}.

It is not possible to obtain accelerated expansion from normal matter
or radiation - as the universe expands, their energy density shrinks
too rapidly, so in order to explain the observations one is forced to
consider other forms of stress energy.  It is very easy to obtain
accelerated expansion from the potential of a self interacting scalar field, although
typically one does not obtain a prolonged period of such behaviour
without fine-tuning which makes it difficult to explain the inflation
required in the early universe.  In the same way it makes it difficult
to arrange a period of acceleration which began only very recently.

It is therfore worth considering other mechanisms which could give rise
to acceleration to see if there are any reasonable alternatives to the
orthodox mechanisms.

In theories with more than 3 spatial dimensions, the values of the
couplings in the 3+1 dimensional theory depend upon the details of the
compactification of the higher dimensions.  Perhaps the simplest
example of this is the ratio between the Newton's constant which
appears in the higher dimensional theory $G_{D}^{-1}=8\pi M_F^{D-2}$
and that which appears in the 3+1 dimensional theory $G_{4}^{-1}=8\pi
M_{Pl}^2$, which is simply the volume of the compact
space\footnote{For a string theory the usual situation is $8\pi
G_{D}=M_F^{2-D}=g_s^{2}l_s^{D-2}/8$ where $g_s$ and $l_s$ are the
string coupling and length respectively \cite{witten}.}.  If one then
allows that volume to vary, the 4D Newton's constant will also vary
relative to the underlying higher dimensional length scale.  

The effect of a varying volume will be to add a dynamical scalar in
front of the 4D Ricci Scalar in the action
${\mathcal{L}}=\sqrt{-g}\Phi R[g]+...$ but it is always possible to
perform a conformal transformation on the metric to the Einstein frame
such that the the action for gravity takes the form (see appendix)

\begin{equation}
S=\frac{M_{Pl}^{2}}{2}\int d^{4}x \sqrt{-g}\left\{R-\frac{1}{2}\partial^\mu\phi \partial_\mu \phi\right\}
\end{equation}
where the scalar $\phi$ represents the volume of the compact space
(variously referred to as the radion, dilaton, modulus, breathing mode
etc.).  If we start in $D$ dimensions and compactify to $3+1$ dimensions $\phi$ is given by
$\cite{argurio}$
\begin{eqnarray}
\phi(t)-\phi_0=-\frac{(D-4)}{2}\sqrt{\frac{D-2}{D-4}}\ln\left(\frac{r(t)}{r_0}\right)&&\nonumber\\
\equiv-\frac{(D-4)}{\beta}\ln\left(\frac{r(t)}{r_0}\right)&&
\label{beta}
\end{eqnarray}
where $r_0$ is the radius of the compact (D-4) torus today so that $\phi-\phi_0$ parameterises the relative change in volume over time. 
The conformal transformation should not affect any physics derived from the
Lagrangian, and indeed one finds that the effect of the varying compact space
has been re-absorbed into a variation of the effective density of matter.  In the Einstein frame the left hand side of the field equations will have the same form as Einstein gravity for given
space-time symmetries. However, the equivalence principle is broken
(the gravitational attraction between two particles will depend on the
local value of the field $\phi$) and the density that one uses to solve the equations will be given by $\rho_{eff}=e^{\beta\phi}\rho$ $\cite{lidsey}$.

If one assumes that the total density is the sum of the matter density and a potential of the (ad-hoc at this stage) form $V(\phi)=\bar{V}e^{-\alpha\phi}$ where $\bar{V}$ and $\alpha$ are constants, the total effective density in the Einstein frame is given by
\begin{equation}
\rho_{total}=\bar{V}e^{-\alpha\phi}+\rho e^{\beta\phi}
\label{mixeddensity}
\end{equation}
and (for $\omega<1/3$) the expectation value of $\phi$ will be at the
minimum of this effective density, i.e.
\begin{equation}
\langle\phi\rangle=\frac{1}{\alpha+\beta}\ln\left(\frac{4}{(1-3\omega)}\frac{\alpha}{\beta}\frac{\bar{V}}{\rho}\right)\qquad
; \qquad \omega < 1/3
\label{expectation}
\end{equation}
so that although $\phi-\phi_0$ is always negative since $r\ge r_0$,
$\phi$ is always positive as we will only be considering situations
where $\rho\ll\bar{V}$.  The expectation value of $\phi$ and the
coupling of the matter to gravity therefore depends upon the local
density of stress-energy. 

The authors of $\cite{chameleon}$ refer to $\phi$ as a chameleon field
when it is behaving like this since the mass of the field changes
according to the density of the local medium (see also \cite{barrow}).  The fact that the expectation value of $\phi$ is a function of the local density is more important for our purposes than its mass.
Consider a universe with a spatially homogeneous distribution of matter.  As the universe expands, the energy density will dilute in the normal way but the expectation value of the field $\phi$
will also change, increasing the coupling of that matter to gravity.
In this way the effective dilution of the gravitating energy due to the expansion
will be reduced and the effective equation of state of the energy will change.
In particular we will be interested in finding out under what conditions this
situation can lead to accelerated expansion.  In order for such an
increase in coupling to compensate for dilution, we require a
potential such as the one written down in equation
(\ref{mixeddensity}) which, in the absence of matter, would cause the
compact space to shrink as the universe expands.  This is not easy to
achieve.

The reader should be aware of some closely related previous work where
the mass of particles changes with the cosmically varying expectation
value of a scalar field $\cite{vamps}$ (see also \cite{rachel}) Here, we are changing the gravitational mass of the particles whilst their inertial mass stays a fixed fraction of $M_{F}$.  
We are considering only cases when the mass scale associated with the energy
density, $\rho^\frac{1}{4}$ is much less than the inverse radius of the
compact space.  It is therefore irrelevant as to whether the matter is
confined to a brane or not, the size of the compact space will not make any
difference to its bare stress energy in the Jordan-string frame since we will assume that any matter which does propagate in the bulk consists only of zero modes.  The low energy gauge coupling of those fields which are able to propagate in the compact space will change with the size of the compact
manifold, but we will assume that no phase transitions occur because of this
effect, and will only consider particles which are given mass via some
Yukawa coupling rather than some confinement scale ('electron' like
particles rather than 'baryon' like ones)
In the next section we will find out how the effective equation of state
behaves for the simple situation presented in equation (\ref{mixeddensity}).
We will then compare the situation with that of an exponential potential and
no matter, in other words power law inflation.  Next we will set up some checks that our scenario must pass in order to be self consistent.
Then we will find out how much expansion it is possible to obtain using matter with such a potential without violating our consistency checks.  After that we will discuss why it is very difficult to find a potential like the one used above but we will show that it is possible to find some well motivated potentials which contain regions where the compact space is dynamically driven to smaller radii.  Finally we will show that it is impossible to obtain acceleration using only the matter and the motion of the scalar field without another potential being present.
\section{Effective equation of state and checks}

We express the relationship between density and pressure with the usual
$\omega$ parameter such that $P=\omega \rho$ and the energy density red-shifts
in the normal way as (see appendix)
\begin{eqnarray}
\rho=\left(\frac{a_0}{a}\right)^{3(1+\omega)}\rho_0\nonumber\\
\dot{\rho}=-3H(1+\omega)\rho
\label{redshift}
\end{eqnarray}
where $H=\dot{a}/a$ is the Hubble expansion factor.  If one then substitute these back into equation ($\ref{mixeddensity}$) one
finds that the effective density can be found analytically
\begin{eqnarray}
\rho_{eff}&=&\left\{\left(\frac{4}{1-3\omega}\frac{\alpha}{\beta}\right)^{\frac{\beta}{\alpha+\beta}}+\left(\frac{4}{1-3\omega}\frac{\alpha}{\beta}\right)^{-\frac{\alpha}{\alpha+\beta}}   \right\}
\bar{V}^{\frac{\beta}{\alpha+\beta}}\rho_0^{\frac{\alpha}{\alpha+\beta}}\left(\frac{a_0}{a}\right)^{\frac{3(1+\omega)\alpha}{\alpha+\beta}}\nonumber\\
&\equiv&\rho_{*}\left(\frac{a_0}{a}\right)^{\frac{3(1+\omega)\alpha}{\alpha+\beta}}
\end{eqnarray}
It becomes clear that
the effective equation of state of the density in the Einstein Frame is given
by the expression
\begin{equation}
\omega_{eff}=\frac{\omega\alpha-\beta}{\alpha+\beta}
\end{equation}
and the two limiting cases show what will happen; if $\alpha\gg\beta$ then the
fluid gravitates in the way that you would expect it to.  If $\beta\gg\alpha$
then the fluid behaves like an inflaton field and the dilution of the matter
is entirely compensated by the increase in its gravitational
coupling.
It is instructive at this point to compare the situation with power law inflation $\cite{power}\cite{lidlyth}$ where there is only an exponential potential and no energy density due to radiation or matter.  The potential therefore has the form
\begin{equation}
V=V_0 e^{-\gamma\psi}
\end{equation}
and the condition for accelerated expansion $V>M_{Pl}^2\dot{\psi}^2$ translates into the inequality
\begin{equation}
\epsilon=\frac{M_{Pl}^2}{2}\left(\frac{1}{V}\frac{\partial V}{\partial \psi}\right)^2=\frac{\gamma^2}{2}\ll 1.
\label{powerlaw}
\end{equation}
We have already seen that things are quite different in our model, and in fact the condition for accelerated expansion, $\omega_{eff}<-1/3$, relates not to the absolute but the relative values of $\alpha$ and $\beta$
\begin{eqnarray}
\omega_{eff}<-\frac{1}{3}&\leftrightarrow& 2\beta>\alpha \quad ({\rm matter}).
\end{eqnarray}
Before we carry on there are some consistency checks which we need to be
fulfilled in order for our analysis to be valid.  

i) The first check is whether or not the kinetic energy $M_{Pl}^2\dot{\phi}^2/2$ is negligible compared to the gravitational energy density of the fluid.  We can find the solution for the scale factor $a$ as a function of time (when $\alpha > 0$)
\begin{equation}
\frac{a(t)}{a_0}=\left[1+\frac{\sqrt{3}}{2}\frac{\alpha(1+\omega)}{\alpha+\beta}\frac{\sqrt{\rho_*}}{M_{Pl}}(t-t_0)\right]^{\frac{2}{3}\frac{\alpha+\beta}{\alpha(1+\omega)}}
\label{afact}
\end{equation}
where we see that as we increase $\beta$, we get closer to
exponential $\omega_{eff}=-1$ behaviour.  Note this is in contrast to the
normal behaviour for the scale factor $a\sim t^{\frac{2}{3(1+\omega)}}$.  Then with equations ($\ref{expectation}$) and ($\ref{redshift}$) we see that we have to have
\begin{equation}
\frac{\rho_{eff}}{\rho_{kinetic}}
=
\frac{2}{3}
\frac{(\alpha+\beta)^4}{\alpha^2(1+\omega)^2}
\gg 1
\label{potdom}
\end{equation}
in order for our assumption that the kinetic energy is negligible to be self-consistent.
ii) The next consistency check is to ensure that the field lies in the minimum of the potential rather than slowly rolling towards that minimum, in other words to ensure that $m^2 \gg H^2$ where $H$ is the Hubble parameter.  It is easy to show that as long as the expansion is dominated by the effective density rather than the kinetic energy of $\phi$
\begin{equation}
\frac{m^2}{H^2}=\frac{3\alpha\beta(\alpha+\beta)(1-3\omega)}{4\alpha+\beta(1-3\omega)}\qquad ; \qquad
m^2=\frac{1}{M_{pl}^2}\frac{\partial^2 V}{\partial\phi^2}
\label{fastroll}
\end{equation}
and we need to ensure that the mass associated with our effective potential is much bigger than the Hubble parameter.
iii) The final consistency check for the time being is one upon the adiabaticity of the contraction of the compact space.  Gravitational waves (or other fields) with momentum in the compact directions show up as massive excitations in the low energy theory known as Kaluza Klein (KK) modes.  The spectrum of these modes is a tower of states of mass $m_{KK}=n/(2\pi r)$ where $n$ runs from 1 upwards.  Because $\phi$, and therefore $r$, is changing in the course of the cosmological evolution in this model, we need to ensure that the time scale over which $r$ changes is always larger than the inverse mass of the lightest KK mode
\begin{eqnarray}
\left|\frac{1}{r(t)}\frac{\partial r(t)}{\partial t}\right|&\ll& m_{KK} \sim \frac{1}{2\pi r(t)}\nonumber\\
\left|2\pi \frac{\partial r(t)}{\partial
t}\right|&=&\frac{\beta}{(D-4)}e^{-\frac{\beta}{D-4}(\phi-\phi_0)}
2\pi r_0 \dot{\phi}\ll 1
\end{eqnarray}
which leads to the expression
\begin{equation}
\frac{\sqrt{3}}{D-4} \frac{\alpha\beta(1+\omega)}{(\alpha+\beta)^2}\left(1+\frac{4}{(1-3\omega)}\frac{\alpha}{\beta}\right)^{1/2}
\frac{2\pi r_0\bar{V}^{\frac{1}{2}}}{M_{Pl}}e^{-\left(\frac{\beta}{D-4}+\frac{\alpha}{2}\right)(\phi-\phi_0)}\ll1.
\label{adiabaticity}
\end{equation}
It is also necessary to make sure that the particles which make up the density are not created as $\phi$ changes. Since the particles follow geodesics in the string frame, it is there where we should consider particle production. Let us imagine that our matter consists of excitations of a scalar field of mass $M$.  If we expand the field into Fourier modes then each mode will obey the equation (see, e.g., \cite{tkachev}) 
\begin{equation}
\frac{d^2h_k}{d\eta^2}+\omega_k^2h_k=0
\end{equation}
where $d\eta=dt/a$, the conformal time. The frequency $\omega_{k}$ is given by 
\begin{equation}
\omega_k^2=k^2+M^2a^2-\frac{1}{a}\frac{d^2a}{d\eta^2}
\end{equation}
so that the effective mass of the mode depends upon the expansion of the space-time in which it is propagating. 

When the effective mass becomes tachyonic, the particle will be produced rapidly, something which we need to avoid if we are to trust our equations. Now transferring back to coordinate time $t$ we find that the effective mass of the mode will become tachyonic when 
\begin{equation}
M^2\ll H^2+\frac{\ddot{a}}{a}
\end{equation}
In the String frame, the mass $M$ will be a constant whereas the value of $M_{Pl}$ will change over time. The scalar will therefore become tachyonic when 
\begin{eqnarray}
M^2&\ll&\frac{2}{3} \frac{\rho}{M_{Pl}^2} +\frac{1}{M_{Pl}}\sqrt{\frac{\rho}{3}}\left\{2\frac{d\ln\rho}{dt}-\frac{d\ln M_{Pl}}{dt} \right\}\nonumber\\
&=&2H^2\left(-2-\frac{d\ln M_{Pl}}{d\ln a}\right)
\end{eqnarray}
and now we can use the results in the rest of the paper to show that 
\begin{equation}
\frac{d\ln M_{Pl}}{d\ln a}=-\frac{\beta}{2}\frac{\dot{\phi}}{H}=-\beta\sqrt{\frac{3\rho_{kin}}{2\rho_{eff}}}=-\frac{3}{2}\frac{\alpha\beta(1+\omega)}{(\alpha+\beta)^2}
\end{equation}
We are considering matter ($\omega=0$) as the source of the energy density in the string frame and we are interested in obtaining acceleration so that $2\beta>\alpha$. With these parameters we find that explosive particle production would only occur when $M^2<-2H^2$, and therefore does not. The reason for this is that while the scale factor is accelerating in the Einstein frame, the expansion in the string frame where the particles propagate is much more gentle. Particles may be produced in small amounts, it would be possible to evolve the mode functions and find the corresponding Boguliubov coeffecients, but the effect will be very small in comparison to the background of particles.

Having listed the consistency checks necessary, we will proceed by considering a toy model so as to find out some typical numbers.
\section{How much expansion?}

There are many free parameters in the model we have outlined above.
In order to consider a particular situation, let us assume that $D=10$
as in a super-symmetric string theory so that $\beta=\sqrt{3}$ and
that $\beta=2\alpha$. 
Putting these values into equation ($\ref{potdom}$) we find that the
potential energy is bigger than the kinetic energy as we require.  The
requirement that the field does not slow roll but stays in its minimum
is fulfilled.  Note that as is often the case when one is dealing with exponential potentials, these two conditions are scale invariant, and if they are fulfilled at some point during the evolution of $\phi$, they will always be true.  However, this scale invariance is broken when one considers the production of KK modes since the ratios between $M_{F}$, $r^{-1}$ and $M_{Pl}$ are all changing over the course of the evolution.  It is therefore the adiabaticity condition which puts a constraint on the number of efolds.
It is well known that around 60 efolds of accelerated expansion in the early universe could solve the horizon and flatness problems \cite{lidlyth}. The number of e-folds in our model is given by
\begin{equation}
N_{efolds}=\int_{t_{start}}^{t_{end}} H
dt=\int_{\phi_{start}}^{\phi_{end}}\frac{1}{a}\frac{da}{d\phi}d\phi
\end{equation}
where $\phi_{start}$ is the start of inflation and we set $\phi_{0}$ as the end of inflation.  One quickly obtains the analytical expression
\begin{equation}
N_{efolds}=\frac{\alpha+\beta}{3(1+\omega)}(\phi_{0}-\phi_{start})
\end{equation}
then since inflation ends when $r=r_0$
\begin{equation}
\frac{r_{start}}{r_0}=\exp\left(\frac{3\beta(1+\omega) N_{efolds}}{(D-4)(\alpha+\beta)}\right)
\end{equation}
which for $D=10$, $2\alpha=\beta=\sqrt{3}$ and $N_{efolds}=60$ requires a factor $10^8$ change for matter.
Let us assume that $\bar{V}=M_{F}^4$, which might be considered natural since $M_F$ is the only scale in our matter sector.  Then using $M_{Pl}^2=M_F^{D-2}(2\pi r_0)^{D-4}$ and our parameters for $\beta$ and $D$, we see that equation (\ref{adiabaticity}) takes the form
\begin{equation}
\frac{1}{9}
\exp\left[{\frac{5}{6}N_{efolds}}\right]\ll
\frac{M_{Pl}}{2\pi r_0 \bar{V}^{\frac{1}{2}}}=\left(\frac{M_{Pl}}{M_{F}}\right)^{\frac{2(D-6)}{(D-4)}}.
\end{equation}

In order to get as much acceleration as possible without the
associated contraction of the compact space resulting in the
non-adiabatic excitation of KK mode quanta, we need to {\sl increase}
the radius of the compact space.  This is simply because in setting
$\bar{V}=M_{F}^4$ we have made the kinetic energy of $\phi$, and
therefore the change in $r$, a stronger function of $M_F$ than
$m_{KK}$ at the beginning of the expansion.

The normal running of the gauge couplings and the non-observation of black holes at particle accelerators forces one to only consider values of $M_{F}$ which are greater than about a TeV.  In this way we are able to constrain the number of e-folds
\begin{equation}
N_{efolds}\le \frac{6}{5}\ln\left[9\left(\frac{10^{18}{\rm GeV}}{10^{3}{\rm GeV}}\right)^{\frac{4}{3}}\right]\simeq 58
\end{equation}
which is coincidentally quite close to the required value.  However,
the fact that we have invented the shape and scale of our potential
rather than using a well motivated one means that the coincidence is
just that.  

However, we are not trying to promote this seriously as a model of
inflation,  partially because we have no motivation for the potential
and partially because it is an incomplete model in as much as it says
nothing about perturbations.  Unlike the inflaton, our field $\phi$ is
not effectively massless, so the normal mechanism for generating
perturbations will be suppressed.  However, there are other ways one
can imagine perturbations being created under such conditions.  One
such mechanism is to invoke the curvaton scenario \cite{curvaton},
namely that there are orthogonal flat directions in the field space in
which iso-curvature perturbations can be produced which later
decay. Another possibility is the recently developed modulon picture, where the coupling of the inflaton to standard model fields is set by the expectation value of a light scalar field \cite{modulon}. In these models, temperature (density) fluctuations in the plasma are created by anisotropies in the decay rate of the inflaton, which is in turn set by fluctuations in the light field responsible for the coupling.

Another obstacle to be wary of is that the radius of the compact torus
would be so large at the beginning of inflation that only very small
values of $\rho$ could be used without the thermal production of $KK$
modes becoming important.  This would lead to a further dissipation of
$\rho$ other than dilution due to the expansion.  However, these KK
modes would also act as sources of gravitational energy which would
give rise to expansion.  The problem is complicated and we leave it
for future studies.

So far we have been dealing with an exponential potential which drives
the compact space towards zero size without discussing its origin.  As
we have mentioned, such a situation is not generic and we shall turn
to that subject now.

\section{Finding an appropriate potential}

We have seen that in order for our model to work and give rise to
expansion, we need to find a potential which pushes us towards small
volumes of the compact space.  A moment's thought makes it clear that
the potential also has to have a positive overall value in order for
the net energy density to be positive at the effective minimum of
$\phi$.  We therefore would like to find a potential which is positive
and decreases as $\phi$ grows. 

The potentials that one often encounters when considering the
stabilisation of higher dimensions in superstring theories come from
non-perturbative effects such as gaugino condensates and instanton
actions.  The classical action $B$ for these instantons generically
takes the form $B\sim g_s^{-1}(R/l_s)^n$ so that their contribution to
the effective potential takes the form $V(\phi)\sim e^{-e^{-\phi}}$
which monotonically increases with $\phi$ $\cite{brustein}$.  This is
discouraging, but precision tests in the solar-system show that we
live in a universe where the majority of gravity is transmitted over
large scale by spin-2 fields rather than scalars.  Some potential must
therefore exist to stabilise and give mass to the $\phi$ field if
there are in reality higher dimensions.

Other kinds of well known effective potentials for the modulus $\phi$ are potentials due to a bare cosmological constant, a non-zero 4-form flux field strength and curvature of the higher dimensional manifold.  The effective potential due to these three contributions is given by \cite{garvil,deruelle}.
\begin{equation}
V(\phi)=M_{Pl}^2\Lambda
e^{\frac{\beta}{2}\phi}+\frac{F_0^2}{2}e^{\frac{3\beta}{2}\phi}-\frac{M_{Pl}^4K}{2}e^{\frac{2}{\beta}\phi}
\end{equation}
where $\Lambda$ is a higher dimensional (geometrical) cosmological
constant, $F_0^2$ parameterises the value of the
4-form field strength and $K$ is negative for a negatively curved compact
space and positive  for a positively curved compact space.  All of these
contributions have the wrong sign in either the exponential or in their
absolute value to provide on their own the kind of potential we are looking
for.  However, if we look at the potential obtained by summing these three contributions we find at some candidate potentials.  
\begin{figure}
\begin{center}
\epsfig{file=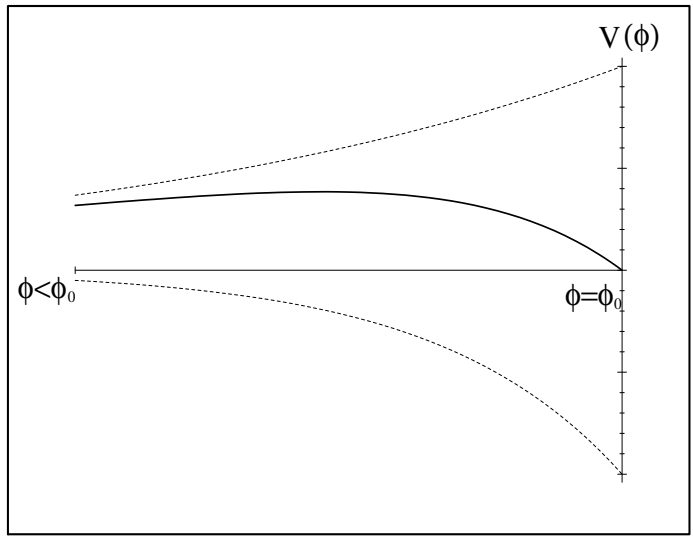,width=8cm,height=10cm,angle=270}
\caption{\label{fin01}\sl Schematic diagram showing how the combined
  contributions of a cosmological constant and a positively curved compact
  space lead to a potential with a region where $\partial V/\partial \phi<0$}
\end{center}
\end{figure}

First of all, it is clear that in order to have any kind of minimum or maximum
we must consider a positively curved compact space so that the curvature contribution to the potential is negative.  The definition of $\beta$ in equation ($\ref{beta}$) makes it clear
that the flux contribution to the potential drops faster than the curvature
term.  These two contributions together can only therefore give rise to a
minimum with a negative absolute value of the potential.  However by combining the cosmological constant term with the curvature term, one can arrive at situations such as that drawn in Figure \ref{fin01} where there is a maximum in the potential and a region where the potential pushes us towards a smaller compact space, although this potential has no minimum.
\begin{figure}
\begin{center}
\epsfig{file=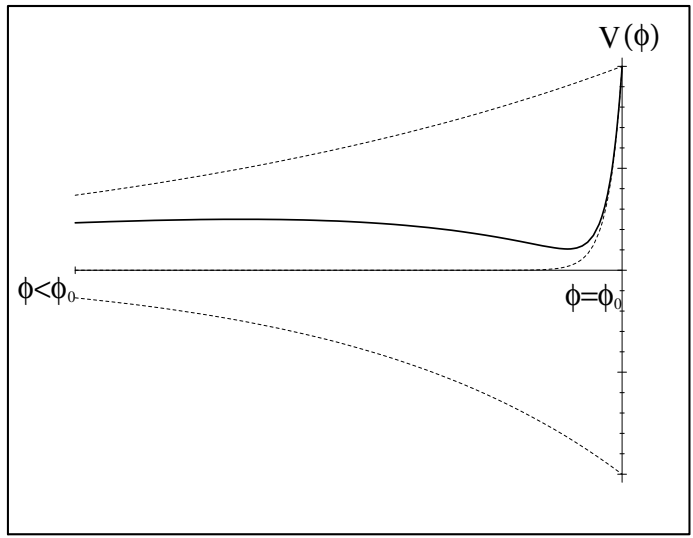,width=8cm,height=10cm,angle=270}
\caption{\label{fin02}\sl Schematic diagram showing how the combined
  contributions of a cosmological constant, non-zero 4-form field strength and
  positively curved compact space lead to a non-trivial potential in the
  Einstein frame}
\end{center}
\end{figure}

Combining all three contributions, one can obtain a well behaved minimum in the way shown in figure ($\ref{fin02}$).  There is a region of this potential where $\partial V/\partial \phi<0$ although it is not clear how natural it would be to tune the density and the size of the compact space so that $\phi$ behaves in the way that we would like.
We will say no more about specific potentials here. 
\section{Acceleration without a potential?}

We have seen that it is difficult, although not impossible, to obtain
potentials which make the compact space shrink over time so might we
therefore obtain acceleration in a different way?  The authors of
\cite{wohlfarth} considered a runaway potential associated with the
breathing mode of a negatively curved compact space, in other words a
positive potential which increases $\sim e^{\frac{2}{\beta}\phi}$ (see
also \cite{ohta}).  Their example was particularly interesting since
they showed that one can obtain acceleration from a purely geometric
source rather than any potential or cosmological constant.  In their
model, they started the field $\phi$ rolling with kinetic energy in
the positive $\phi$ direction so as to obtain a brief period of
accelerated expansion at the point at which $\phi$ climbs the
potential, stops and starts to roll back down again.  Note that
equation ($\ref{powerlaw}$) shows us that this curvature potential is
too steep to give rise to inflation, although there are variations on
this theme which can give rise to more acceleration \cite{more}.

In our model, energy density acts as an effective exponential
potential for $\phi$, so it would be interesting to see if we can
obtain a period of acceleration in an analogous way, i.e. without any
potential or cosmological constant.  Before we begin we can see
straight away that we face the added difficulty of overcoming the
dilution of the energy density as the universe expands, in other
words, as our field $\phi$ runs up the slope, the absolute value of
the slope is decreasing as our matter becomes more
dilute. 

For the case in reference \cite{wohlfarth} the expression for the acceleration in the Einstein frame is given by the normal one for a scalar field moving in a potential
\begin{equation}
\frac{\ddot{a}}{a}=-\frac{1}{6M_{Pl}^2}(\rho+3P)=-\frac{1}{3}(\dot{\phi}-\frac{V(\phi)}{M_{Pl}^2})
\end{equation}
and one can see that as $\phi$ comes to a stop, the universe will
accelerate.

It is almost as easy to obtain the corresponding expression for our situation. Using the equation of motion for the scalar field
\begin{equation}
\ddot{\phi}+3H\dot{\phi}+\frac{1}{4}(1-3\omega)\frac{\beta\rho}{M_{Pl}^2}e^{\beta\phi}=0
\end{equation}
and the expression for the hubble constant $H$
\begin{equation}
H^2=\frac{1}{3}\left(\frac{\rho}{M_{Pl}^2} e^{\beta\phi}+\frac{1}{2}\dot{\phi}^2\right)
\end{equation}
as well as the expression for the dilution of $\rho$ equation ($\ref{redshift}$) we find that
\begin{equation}
\frac{\ddot{a}}{a}=\dot{H}+H^2=-\frac{1}{6}(1-3\omega)\frac{\rho e^{\beta\phi}}{M_{Pl}^2}-\frac{\dot{\phi^2}}{3}+\frac{1+\omega}{8}\frac{\rho e^{\beta\phi}}{M_{Pl}^2}\frac{\beta\dot{\phi}}{H}
\end{equation}
which is always less than zero and acceleration will not occur.  This is not so surprising, the
brief period of acceleration which occurs in the model described in
reference \cite{wohlfarth} when $\phi$ is stationary corresponds, in
our model, to a brief period of time where $\phi$ is not changing and
the matter is therefore red-shifting as we would normally
expect it to.  It is therefore not possible to obtain any acceleration
without the presence of some potential.

\section{Discussion}
In this paper we showed that in models with compactified higher
dimensions, changes in the size of the compact space change the
effective equation of state of matter.  In particular,
in the presence of a potential which tends to reduce the size of the
compact space, one can obtain acceleration from normal matter.

Assuming a simple form for this potential we have calculated the
maximum amount of expansion one could get from a such a model without
non-adiabatically producing KK excitations around the compact space.
If we did excite KK modes, their effect on the evolution of the
compact and non-compact spaces would be extremely difficult to
calculate.  For a prolonged period of acceleration, the compact space
would have to start out very large compared to its size at the end of
this period.

We have also tried to explain why it is difficult to find potentials
of a suitable form to lead to acceleration, although we have shown
that realistic potentials already exist which contain regions where
the compact space would be pushed to smaller sizes as the matter or
radiation becomes diluted.  We ended by pointing out that it is
hopeless to try and get acceleration without the presence of a
potential.

So far we have said nothing about Dark energy.  It has been pointed
out recently that if one is able to find an appropriate potential of
the form that we have used in this paper one can give a mass to $\phi$
which changes according to the local density $\cite{chameleon}$.  In
particular, tests of tensor gravity could be passed in denser mediums
such as the solar system or neutron star binary systems, while a
scalar-tensor theory would describe gravity in the low density voids
between clusters of galaxies.  The equation of state of dark matter,
could be reduced in these regions as it
is in other models of varying mass particles (VAMPS)
\cite{vamps}\cite{rachel}.  In this way, one might hope to shed light
on the coincidence between the energy density of dark energy and dark matter.

\section*{Appendix}
We assume a compact toroidal space where all of the higher dimensions
have the same size.  We call the 4+D dimensional metric $G_{MN}$ and
since it is diagonal we can easily split it into a 4D part
$g_{\mu\nu}$ and a higher dimensional part $h_{ij}$ with components
$h_{ii}=h^{\frac{1}{D-4}}$ where $h$ is the determinant of $h_{ij}$.

\begin{equation}
S=\frac{M_{Pl}^{2}}{2}\int d^{4}x \sqrt{-g}\sqrt{\frac{h}{h_0}}\left\{R[g]+\frac{(D-5)}{(D-4)}\frac{1}{4}g^{\mu\nu}\partial_\mu \ln h \partial_\nu \ln h\right\}
\end{equation}
Here, $x$ are the coordinates of the 4-dimensional space and  the four
dimensional Planck mass $M_{Pl}$ is given by
\begin{equation}
M_{Pl}^2=(8\pi G)^{-1}=M_F^{D-2}\int d^{D-4}y \sqrt{-h_0} =M_F^{D-2}(2\pi r)^{D-4}
\end{equation}
where $y$ are the coordinates and $r$ are the radii of the tori of the compact space.
Now a conformal transformation of the metric of the form $g_{\mu\nu}=e^{2\sigma}\tilde{g}_{\mu\nu}$ allows us to go to the Einstein frame and define a canonically normalised scalar field $\phi$ 
\begin{equation}
\phi-\phi_0=-\frac{1}{4}a\ln \left(\frac{h}{h_0}\right)=a\sigma\qquad ; \qquad a=2\sqrt{\frac{D-2}{D-4}}
\end{equation}
such that
\begin{equation}
S=\frac{M_{Pl}^{2}}{2}\int d^{4}x \sqrt{-\tilde{g}}\left\{\tilde{R}-\frac{1}{2}\partial^\mu \phi \partial_\mu \phi\right\}.
\end{equation}
The stress energy tensor of matter in the string frame has coupling in the Einstein frame given by \cite{lidsey}
\begin{equation}
\tilde{T}^{\mu\nu}=e^{6\sigma}T^{\mu\nu}
\end{equation}
and the new Christoffel symbol after the transformation is given by
\begin{equation}
\tilde{\Gamma}^{\gamma}_{\mu\nu}={\Gamma}^{\gamma}_{\mu\nu}-{g}^{\gamma}_{\mu}\sigma_{,\nu}-{g}^{\gamma}_{\nu}\sigma_{,\mu}+{g}_{\mu\nu}{g}^{\gamma\kappa}\sigma_{,\kappa}
\end{equation}
so that we can rewrite the energy conservation equation:
\begin{equation}
\tilde{\nabla}_{\beta}T^{\alpha\beta}={\nabla}_{\beta}{T}^{\alpha\beta}-6T^{\alpha\gamma}\sigma,{\gamma}+{T}\sigma^{,\alpha}
\end{equation}
The covariant derivative of the stress energy tensor of the matter in the Einstein frame with the $\sigma$ dependant coupling is
\begin{eqnarray}
\tilde{\nabla}_{\beta}\left(e^{6\sigma}T^{\alpha\beta}\right)&=&T^{\alpha\beta}\tilde{\nabla}_{\beta}e^{6\sigma}+e^{6\sigma}\tilde{\nabla}_{\beta}T^{\alpha\beta}\nonumber\\
&=&e^{6\sigma}{\nabla}_{\beta}T^{\alpha\beta}+e^{6\sigma}g^{\alpha\beta}T
\sigma_{,\beta}\nonumber\\
\alpha=0\Rightarrow&=&e^{6\sigma}\left(\dot{\rho}+3H(1+\omega)\rho\right)+e^{4\sigma}\left(1-3\omega\right)\rho\dot{\sigma}
\label{}
\end{eqnarray}

Next we write $\tilde{T}^{\beta\alpha}_{\phi}$ which is the stress energy tensor of the scalar field defined in the normal way in the Einstein frame as
\begin{equation}
\tilde{T}^{\alpha\beta}_{\phi}=M_{Pl}^2\partial^{\alpha}\phi\partial^{\beta}\phi-\frac{1}{2}g^{\alpha\beta}\left(M_{Pl}^2\partial^{\mu}\phi\partial_{\mu}\phi+2\bar{V}e^{-\alpha\phi}\right)
\end{equation}
where the potential is the one used in the main body of the text.  It is the covariant derivative of the sum of all the stress energies which must be zero to match the Bianchi identities so that
\begin{equation}
\tilde{\nabla}_{\alpha}\left(e^{6\sigma}{{T}_{M}}^{\beta\alpha}+{\tilde{T}_{\phi}}^{\beta\alpha}\right)=0
\end{equation}
The relationship between the the canonically normalised field
$\phi$ and the field $\sigma$ used in the conformal transformation is
$4\sigma=\beta\phi$. Then the above equation splits into a
conservation equation for the matter and an equation of motion for $\phi$
\begin{eqnarray}
&&\dot{\rho}+3H(1+\omega)\rho=0 \nonumber\\
&&\dot{\phi}\left(M_{pl}^2\ddot{\phi}-\alpha\bar{V}^{-\alpha\phi}+\frac{\beta}{4}(1-3\omega)\rho e^{\beta\phi}+3M_{pl}^2H\dot{\phi}\right)=0
\end{eqnarray}
which shows that there is an effective potential for the $\phi$ field of the form
\begin{equation}
V_{eff}=\bar{V}e^{-\alpha\phi}+\frac{1}{4}\rho(1-3\omega) e^{\beta\phi}
\end{equation}
as used in the text.
\section*{Acknowledgments}
I am extremely grateful for conversations with Tom Dent, Raf Guedens, Jim
Lidsey, Laura Lopez-Honorez and Michel Tytgat but any errors are entirely due to the author.
I am also grateful for financial support from an IISN grant and the IUAP program of the Belgian Federal Government.


\begin{thebibliography}{99}

\bibitem{sn1a}
R.~A.~Knop {\it et al.},
[arXiv:astro-ph/0309368].
\bibitem{witten}
E.~Witten,
Nucl.\ Phys.\ B {\bf 471}, 135 (1996)
[arXiv:hep-th/9602070].
\bibitem{argurio}
R.~Argurio,
``Brane physics in M-theory,''
[arXiv:hep-th/9807171]
\bibitem{lidsey}
J.~E.~Lidsey, D.~Wands and E.~J.~Copeland,
Phys.\ Rept.\  {\bf 337}, 343 (2000)
[arXiv:hep-th/9909061].
\bibitem{chameleon}
J.~Khoury and A.~Weltman, 
[arXiv:astro-ph/0309300], 
[arXiv:astro-ph/0309411]
\bibitem{barrow}
J.~D.~Barrow and C.~O'Toole,
[arXiv:astro-ph/9904116],
J.~D.~Barrow and D.~F.~Mota,

Class.\ Quant.\ Grav.\  {\bf 20}, 2045 (2003)
[arXiv:gr-qc/0212032].
\bibitem{vamps}
J.~A.~Casas, J.~Garcia-Bellido and M.~Quiros,
Class.\ Quant.\ Grav.\  {\bf 9}, 1371 (1992)
[arXiv:hep-ph/9204213].
J.~Garcia-Bellido,
Int.\ J.\ Mod.\ Phys.\ D {\bf 2}, 85 (1993)
[arXiv:hep-ph/9205216].
G.~W.~Anderson and S.~M.~Carroll,
[arXiv:astro-ph/9711288].
G.~R.~Farrar and P.~J.~E.~Peebles,
arXiv:astro-ph/0307316.
R.~Fardon, A.~E.~Nelson and N.~Weiner,
[arXiv:astro-ph/0309800].
\bibitem{rachel}
L.~Amendola,
Phys.\ Rev.\ D {\bf 60}, 043501 (1999)
[arXiv:astro-ph/9904120].
R.~Bean,
Phys.\ Rev.\ D {\bf 64}, 123516 (2001)
[arXiv:astro-ph/0104464].
\bibitem{power}
L.~F.~Abbott and M.~B.~Wise,
Nucl.\ Phys.\ B {\bf 244}, 541 (1984),
F.~Lucchin and S.~Matarrese,
Phys.\ Rev.\ D {\bf 32}, 1316 (1985).
\bibitem{lidlyth}
A.~R.~Liddle and D.~H.~Lyth,
Phys.\ Rept.\  {\bf 231}, 1 (1993)
[arXiv:astro-ph/9303019].
\bibitem{tkachev}
G.~F.~Giudice, I.~Tkachev and A.~Riotto,
JHEP {\bf 9908}, 009 (1999)
[arXiv:hep-ph/9907510].

\bibitem{curvaton}
D.~H.~Lyth and D.~Wands,
Phys.\ Lett.\ B {\bf 524}, 5 (2002)
[arXiv:hep-ph/0110002].
T.~Moroi and T.~Takahashi,
Phys.\ Lett.\ B {\bf 522}, 215 (2001)
[Erratum-ibid.\ B {\bf 539}, 303 (2002)]
[arXiv:hep-ph/0110096].
\bibitem{modulon}
G.~Dvali, A.~Gruzinov and M.~Zaldarriaga,
Phys.\ Rev.\ D {\bf 69}, 023505 (2004)
[arXiv:astro-ph/0303591].
L.~Kofman,
arXiv:astro-ph/0303614.

\bibitem{brustein}
R.~Brustein, S.~P.~de Alwis and E.~G.~Novak,
Phys.\ Rev.\ D {\bf 68}, 043507 (2003)
[arXiv:hep-th/0212344].
\bibitem{deruelle}
N.~Deruelle, J.~Garriga and E.~Verdaguer,
Phys.\ Rev.\ D {\bf 43}, 1032 (1991).
\bibitem{garvil}
J.~Garriga and A.~Vilenkin,
Phys.\ Rev.\ D {\bf 64}, 023517 (2001)
[arXiv:hep-th/0011262].

\bibitem{wohlfarth}
P.~K.~Townsend and M.~N.~R.~Wohlfarth,
Phys.\ Rev.\ Lett.\  {\bf 91}, 061302 (2003)
[arXiv:hep-th/0303097].
\bibitem{ohta}
N.~Ohta,
Phys.\ Lett.\ B {\bf 558}, 213 (2003)
[arXiv:hep-th/0301095].
\bibitem{more}
C.~M.~Chen, P.~M.~Ho, I.~P.~Neupane, N.~Ohta and J.~E.~Wang,
JHEP {\bf 0310}, 058 (2003)
[arXiv:hep-th/0306291].
I.~P.~Neupane,
\end{thebibliography}
\end{document}